# Universality and the approach to the continuum limit in lattice gauge theory

ALPHA Collaboration

Giulia de Divitiis[a], Roberto Frezzotti[a], Marco Guagnelli[a], Martin Lüscher[b], Roberto Petronzio[a], Rainer Sommer[c], Peter Weisz[d] and Ulli Wolff[e]

[a] Dipartimento di Fisica, Università *Tor Vergata*, and INFN, Sezione di Roma II, Viale della Ricerca Scientifica, I-00133 Roma, Italy
[b] Deutsches Elektronen-Synchrotron DESY, Notkestrasse 85, D-22603 Hamburg, Germany
[c] CERN, Theory Division, CH-1211 Geneva 23, Switzerland
[d] Max-Planck-Institut für Physik, Föhringer Ring 6, D-80805, Munich, Germany
[e] Humboldt Universität, Institut für Physik, Invalidenstrasse 110, D-10099 Berlin, Germany

**Abstract**

The universality of the continuum limit and the applicability of renormalized perturbation theory are tested in the SU(2) lattice gauge theory by computing two different non-perturbatively defined running couplings over a large range of energies. The lattice data (which were generated on the powerful APE computers at Rome II and DESY) are extrapolated to the continuum limit by simulating sequences of lattices with decreasing spacings. Our results confirm the expected universality at all energies to a precision of a few percent. We find, however, that perturbation theory must be used with care when matching different renormalized couplings at high energies.

## 1. Introduction

The continuum limit of lattice QCD is known to exist to all orders of perturbation theory [1]. Up to finite renormalizations the limit is independent of the details of the lattice formulation and the resulting Feynman amplitudes agree with those computed via dimensional regularization.

At the non-perturbative level the existence and universality of the continuum limit still needs to be established. One would also be interested to know whether renormalized perturbation theory is indeed an asymptotically correct expansion of the full amplitudes at high energies. These questions must eventually be decided by rigorous analyses [2], but until this very difficult task is mastered, some insight can be gained through numerical simulations. This is not trivial either, because one needs rather precise data to be able to observe a variation of the chosen physical quantities in the accessible range of lattice spacings. A further difficulty is that renormalization and cutoff effects must be carefully disentangled from each other.

A particularly suitable quantity to consider in this context is the running coupling $\alpha(q)$, because a large range of momenta $q$ can be probed in this case by applying the finite-size scaling technique of ref.[3]. Computations of $\alpha(q)$ along these lines have previously been performed in the pure SU(2) and SU(3) Yang-Mills theories [4–9]. Taking advantage of the massive computer power provided by the APE machines at Rome II and DESY, we have now been able to extend these calculations in several directions, and it is our aim here to discuss the results of this effort.

The model considered is the pure SU(2) gauge theory, the simplest case where the questions mentioned above can be addressed. Compared to our earlier work on this theory, we have simulated larger lattices (thus moving closer to the continuum limit) and generated significantly larger ensembles of field configurations. In some cases a Symanzik improved lattice action has been employed. As a result one is able to control the continuum limit at a higher level of precision and with greater confidence. Additional opportunities to study the universality of the continuum limit and the applicability of renormalized perturbation theory are obtained by comparing two independent running couplings, with significantly different low-energy behaviour and different cutoff dependence.

We assume the reader is familiar with the finite-size technology explained in detail in refs.[3–9]. Our notations and some information on the numerical simulations are collected in sects. 2 and 3. The extrapolation of the numeri-cal data to the continuum limit is discussed in sect. 4. We then address the question of how well perturbation theory describes the evolution of the running couplings and the relation between them (sect. 5). A discussion of the applicability of bare perturbation theory is also included here. After that we are well prepared to compute the running coupling $\alpha_{\overline{\rm MS}}(q)$ in the $\overline{\rm MS}$ scheme of dimensional regularization at large momenta $q$ given in units of a suitable low-energy scale (sect. 6). Conclusions are drawn in sect. 7 and a compilation of simulation data is included in appendix A.

## 2. Running couplings

In this paper we shall discuss two running couplings, $\alpha_{\rm SF}(q)$ and $\alpha_{\rm TP}(q)$. The subscripts SF and TP stand for "Schrödinger functional" and "twisted Polyakov loop", respectively, indicating the physical amplitudes from which the couplings are extracted. They have already been introduced in our previous papers on the subject [4–9] so that here we only briefly recall their definitions. For simplicity they are given in the language of the continuum theory, but it is straightforward to pass to the lattice formulation (cf. sect. 3).

### *2.1 Preliminaries*

As indicated in sect. 1 attention is restricted to the pure SU(2) Yang-Mills theory in four dimensions. The gauge potential $A_\mu(x)$, $\mu = 0,\ldots,3$, thus takes values in the Lie algebra of SU(2), which we choose to be the linear space of all traceless anti-hermitean $2\times 2$ matrices. Space-time is assumed to be a hyper-cubical box of size $L$ with boundary conditions to be specified below. The action is given by

$$S[A] = -\frac{1}{2g_0^2}\int_0^L {\rm d}^4x\,{\rm tr}\,\{F_{\mu\nu}F_{\mu\nu}\}, \qquad (2.1)$$

where $F_{\mu\nu}(x)$ denotes the field tensor,

$$F_{\mu\nu} = \partial_\mu A_\nu - \partial_\nu A_\mu + [A_\mu, A_\nu], \qquad (2.2)$$

and $g_0$ the bare gauge coupling.

The basic idea of the finite-size method of ref.[3] is to study the properties of the system as a function of the box size $L$. In particular, the couplings $\alpha_{\rm SF}(q)$ and $\alpha_{\rm TP}(q)$ are chosen to depend on $L$ but on no other external scale. They are, therefore, running with $L$ and the momentum $q$ is accordingly given by

$$q = 1/L. \tag{2.3}$$

There are many different ways to introduce such couplings. Our choices were made to meet a number of practical criteria (cf. ref.[6]).

For any particular running coupling $\alpha(q)$ we define the associated coupling $\bar{g}^2(L) = 4\pi\alpha(q)$ and the Callan-Symanzik $\beta$–function

$$\beta(\bar{g}) = -L\frac{\partial\bar{g}}{\partial L}. \tag{2.4}$$

In perturbation theory we have

$$\beta(\bar{g}) \underset{\bar{g}\to 0}{\sim} -\bar{g}^3 \sum_{n=0}^{\infty} b_n \bar{g}^{2n}, \tag{2.5}$$

with $b_0 = 11/24\pi^2$ and $b_1 = 17/96\pi^4$ being the usual universal coefficients. All other (three loop and higher order) coefficients depend on the chosen scheme.

Another quantity of interest is the step scaling function $\sigma(s, u)$ which describes the evolution of the running coupling under changes of $L$ by a factor $s$ [3]. Explicitly, if the coupling is equal to $u$ at scale $L$, its value at scale $sL$ is determined by eq.(2.4) and we may define $\sigma(s, u) = \bar{g}^2(sL)$. Since $\sigma(s, u)$ is just an integrated form of the $\beta$–function it is scheme dependent too.

*2.2 Definition of* $\alpha_{\rm SF}$

In this case we impose periodic boundary conditions in the space directions,

$$A_\mu(x + L\hat{k}) = A_\mu(x), \tag{2.6}$$

and require

$$A_k(x) = \begin{cases} C_k(\mathbf{x}) & \text{at } x_0 = 0, \\ C'_k(\mathbf{x}) & \text{at } x_0 = L. \end{cases} \tag{2.7}$$

The index $k$ runs from 1 to 3 and $\hat{k}$ denotes the unit vector in direction $k$. For the classical boundary fields $C$ and $C'$ we take

$$C_k(\mathbf{x}) = \eta\tau_3/iL,$$

$$C_k'(\mathbf{x}) = (\pi - \eta)\tau_3/iL, \tag{2.8}$$

where $0 < \eta < \pi$ is an adjustable parameter and $\tau_3$ the third Pauli matrix.

The Schrödinger functional,

$$\mathcal{Z} = \int \mathrm{D}[A]\, \mathrm{e}^{-S[A]}, \tag{2.9}$$

involves an integration over all gauge fields $A$ with fixed boundary values $C$ and $C'$. It may be interpreted as the (euclidean) propagation kernel for going from the initial gauge field configuration $C$ at time $x_0 = 0$ to the final configuration $C'$ at time $x_0 = L$. At small couplings it is dominated by the field configuration $B$ with least action. For the boundary fields specified above we have

$$\begin{aligned} B_0(x) &= 0, \\ B_k(x) &= [x_0 C_k' + (L - x_0)C_k]/L, \end{aligned} \tag{2.10}$$

and the effective action $\Gamma = -\ln \mathcal{Z}$ may be expanded in the series

$$\Gamma = g_0^{-2}\Gamma_0 + \Gamma_1 + g_0^2\Gamma_2 + \ldots \tag{2.11}$$

with $\Gamma_0 = g_0^2 S[B]$.

Power counting and an explicit one-loop calculation suggest that the effective action is renormalizable up to a divergent additive constant. So if we differentiate with respect to the boundary fields,

$$\Gamma' = \frac{\partial \Gamma}{\partial \eta}, \tag{2.12}$$

the divergent part is removed and a renormalized coupling $\bar{g}_{\mathrm{SF}}$ may be defined through

$$\bar{g}_{\mathrm{SF}}^2 = \Gamma_0'/\Gamma'|_{\eta=\pi/4}\,. \tag{2.13}$$

Note that $L$ is the only external scale on which $\bar{g}_{\mathrm{SF}}$ depends. To one-loop order its relation to the running coupling in the $\overline{\mathrm{MS}}$ scheme of dimensional regularization is ($\alpha_{\mathrm{SF}} = \bar{g}_{\mathrm{SF}}^2/4\pi$)

$$\alpha_{\overline{\mathrm{MS}}} = \alpha_{\mathrm{SF}} + 0.9433 \times (\alpha_{\mathrm{SF}})^2 + \ldots \tag{2.14}$$

where both couplings are evaluated at the same momentum $q$. The relation to the bare lattice coupling is now known to two loops [18] (cf. sect. 5).

### *2.3 Definition of* $\alpha_{\rm TP}$

The definition of $\alpha_{\rm TP}$ involves correlation functions of the Polyakov loops

$$
\begin{aligned}
P_1(x_0,x_2,x_3) &= \ {\rm Tr}\ \left\{ P{\rm e}^{\int_0^L {\rm d}x_1 A_1(x)}\Omega_1 \right\} {\rm e}^{-i\pi x_2/L}, \\
P_3(x_0,x_1,x_2) &= \ {\rm Tr}\ \left\{ P{\rm e}^{\int_0^L {\rm d}x_3 A_3(x)} \right\},
\end{aligned}
\tag{2.15}
$$

with twisted periodic boundary conditions on the gauge potentials, viz.

$$
A_\mu(x+L\hat\nu) = \Omega_\nu A_\mu(x)\Omega_\nu^\dagger. \tag{2.16}
$$

The unitary "twist matrices" $\Omega_\nu$ satisfy

$$
\Omega_0 = \Omega_3 = 1, \qquad \Omega_1\Omega_2 = -\Omega_2\Omega_1, \tag{2.17}
$$

and the extra phase in the definition of $P_1$ is included to guarantee that this observable is a strictly periodic function of the coordinates $x_0$, $x_2$ and $x_3$.

Twisted boundary conditions eliminate degenerate toron configurations and hence enable standard perturbative computations. The perturbation expansion of the two-point correlation function of $P_1$ starts at order $g_0^2$, while the correlation function of $P_3$ has a connected part of order 1. We may thus define a running coupling $\bar g^2_{\rm TP}(L)$ through

$$
\bar g^2_{\rm TP}(L) = k\,\frac{\int {\rm d}x_2{\rm d}x_3\,\langle P_1(L/2,x_2,x_3)P_1(0,0,0)^*\rangle}{\int {\rm d}x_1{\rm d}x_2\,\langle P_3(L/2,x_1,x_2)P_3(0,0,0)^*\rangle}, \tag{2.18}
$$

where the normalization constant $k = 14.459\ldots$ is chosen such that $\bar g^2_{\rm TP}(L) = g_0^2$ to lowest order of perturbation theory. Linear divergences affecting the numerator and the denominator factorize [10] and cancel in the ratio (2.18) which is, therefore, renormalizable and well-defined in the continuum limit. This has been shown to be correct in an explicit one-loop calculation [7], which also yields the relation

$$
\alpha_{\overline{\rm MS}} = \alpha_{\rm TP} - 0.5584\times(\alpha_{\rm TP})^2 + \ldots \tag{2.19}
$$

between $\alpha_{\rm TP}(q) = \bar{g}^2_{\rm TP}(L)/4\pi$ and the coupling in the $\overline{\rm MS}$ scheme of dimensional regularization.

At large $L$ the choice of boundary conditions becomes inessential and the correlations of Polyakov loops in twisted and ordinary directions tend to the same value. We thus expect that $\alpha_{\rm TP}(q)$ converges to $k/4\pi = 1.151\ldots$ when $q \to 0$. The infrared behaviours of $\alpha_{\rm TP}$ and $\alpha_{\rm SF}$ are thus completely different (the latter diverges exponentially [3]).

## 3. Lattice formulation and numerical simulation

### *3.1. Lattice definition of* $\alpha_{\rm SF}$ *and* $\alpha_{\rm TP}$

We consider a hypercubic lattice with spacing $a$ and size $L^4$. The lattice gauge field is denoted by $U(x,\mu)$. In the case of the Schrödinger functional we impose periodic boundary conditions in the spatial directions, while at time $x_0 = 0$ and $x_0 = L$ we require

$$U(x,k)|_{x_0=0} = \exp\{aC_k\}, \qquad U(x,k)|_{x_0=L} = \exp\{aC'_k\}, \tag{3.1}$$

where $C_k$ and $C'_k$ are given by eq.(2.8). The lattice action

$$S[U] = \frac{1}{g_0^2}\sum_p w(p)\,{\rm tr}\,\{1 - U(p)\} \tag{3.2}$$

involves a summation over all oriented plaquettes $p$ on the lattice with $U(p)$ being the parallel transporter around $p$. The weight $w(p)$ is given by

$$w(p) = \begin{cases} c_t(g_0) & \text{if } p \text{ is a time-like plaquette touching the boundary,} \\ 1 & \text{elsewhere.} \end{cases} \tag{3.3}$$

Note that in the present context we do not need to specify the weight of the spatial plaquettes at $x_0 = 0$ and $x_0 = L$ since their contribution vanishes for the boundary values chosen.

As discussed in ref.[4] the lattice boundaries give rise to O($a$) lattice artifacts which can, in principle, be cancelled by an appropriate choice of $c_t(g_0)$. In perturbation theory we have

$$c_t(g_0) = 1 + c_t^{(1)} g_0^2 + c_t^{(2)} g_0^4 + \ldots \tag{3.4}$$

Setting $c_t(g_0) = 1$ is referred to as tree-level improvement in this paper. For 1-loop improvement we include [4]

$$c_t^{(1)} = -0.0543,$$

and for 2-loop improvement [18]

$$c_t^{(2)} = -0.0115$$

is taken in addition.

The Schrödinger functional and the effective action $\Gamma$ on the lattice are defined through

$$\mathcal{Z} = \exp\{-\Gamma\} = \int \mathrm{D}[U]\, \mathrm{e}^{-S[U]}, \qquad \mathrm{D}[U] = \prod_{x,\mu} \mathrm{d}U(x,\mu). \tag{3.5}$$

At small couplings the integral is dominated by (the gauge orbit of) the classical field

$$V(x,\mu) = \exp\{aB_\mu(x)\}$$

with $B_\mu(x)$ given by eq.(2.10). The corresponding action is

$$\Gamma_0 = 24\frac{L^4}{a^4}\sin^2\left\{\frac{(\pi-2\eta)a^2}{2L^2}\right\}. \tag{3.6}$$

The running coupling $\bar{g}^2_{\mathrm{SF}}(L)$ is now defined by eq.(2.13) with $\eta$-derivatives of the lattice quantities on the right hand side.

In numerical simulations the coupling is obtained through

$$\bar{g}^2_{\mathrm{SF}}(L) = \Gamma_0' \left\langle \frac{\partial S}{\partial \eta} \right\rangle^{-1}. \tag{3.7}$$

Explicitly the "observable" $\partial S/\partial\eta$ is given by

$$\frac{\partial S}{\partial \eta} = -c_t(g_0)\frac{2a^3}{g_0^2 L}\sum_{\mathbf{x}}\sum_{k=1}^{3}\left\{E_k'(\mathbf{x}) + E_k(\mathbf{x})\right\}, \tag{3.8}$$

where

$$E_k(\mathbf{x}) = \frac{1}{ia^2}\mathrm{tr}\left\{\tau_3 V(x,k)U(x+a\hat{k},0)U(x+a\hat{0},k)^{-1}U(x,0)^{-1}\right\}_{x^0=0} \tag{3.9}$$

($E_k'$ is defined analogously at $x_0 = L$).

In the case of the coupling $\bar{g}^2_{\rm TP}(L)$ we impose twisted periodic boundary conditions, i.e. we initially require $U(x + L\hat{\nu}, \mu) = \Omega_\nu U(x,\mu)\Omega_\nu^\dagger$. By a change of variables the theory is then rewritten as a theory with ordinary periodic boundary conditions, but with a modified Wilson action where the sign of some plaquettes is reversed (see e.g. ref.[11]). In this formulation the Polyakov loops are given by

$$\begin{aligned} P_1(x_0,x_2,x_3) &= \mathrm{Tr}\left\{\prod_{x_1=0}^{L-a} U(x,1)\right\} \mathrm{e}^{-i\pi x_2/L}, \\ P_3(x_0,x_1,x_2) &= \mathrm{Tr}\left\{\prod_{x_3=0}^{L-a} U(x,3)\right\}, \end{aligned} \qquad (3.10)$$

and $\bar{g}^2_{\rm TP}(L)$ is now again defined through eq.(2.18), where the integrals are to be replaced by the corresponding sums over lattice points.

It should be noted at this point that the constant $k$ is taken to be the same as in the continuum theory. On the lattice $\bar{g}^2_{\rm TP}(L)$ is, therefore, not exactly equal to $g_0^2$ at tree level of perturbation theory. The coupling is still proportional to $g_0^2$ with a proportionality constant differing from 1 by terms of order $a^2$.

To reduce statistical fluctuations in numerical simulations we made use of the 1-link integral technique which replaces each link in the Polyakov loop by its average value in the external field produced by the neighbouring links [12]. If $\overline{P}_k$ denotes the Polyakov loop $P_k$ averaged in this way, the coupling $\bar{g}^2_{\rm TP}(L)$ was thus obtained by computing the expectation values of the observables

$$\begin{aligned} \mathcal{O}_1 &= a^2 \sum_{x_2,x_3} \overline{P}_1(L/2,x_2,x_3)\overline{P}_1(0,0,0)^*, \\ \mathcal{O}_3 &= a^2 \sum_{x_1,x_2} \overline{P}_3(L/2,x_1,x_2)\overline{P}_3(0,0,0)^*, \end{aligned} \qquad (3.11)$$

and taking their ratio.

*3.2 Simulation algorithm*

As already mentioned, the numerical simulations have been performed on the APE computers at DESY and Rome II. In the most commonly used configuration these computers consist of a mesh of 128 computing nodes. The local memory is 1 Mword so that lattices with $L/a \leq 24$ can be simulated on subsets of 8 nodes, i.e. in this case there are 16 independent lattices on the machine. Larger lattices must be distributed over more nodes and the performance of the simulation algorithm is then slightly reduced due to the communication losses.

In all cases a hybrid over-relaxation algorithm was employed where $N_{\rm OR}$ exactly microcanonical over-relaxation sweeps are followed by 1 Fabricius-Haan [13,14] heatbath sweep. The high-quality random number generator of ref.[15] was used. $N_{\rm OR}$ should be chosen so as to minimize the autocorrelation times of the quantities of interest. We have not tried to optimize this parameter, but found, after some testing, that $N_{\rm OR} = L/2a$ is in general a good choice.

We now need to be a bit a more specific on the order in which the link variables are updated, because it turns out (to our surprise) that the autocorrelation times depend on these seemingly marginal details (cf. subsect. 3.3). Only the lattices which fit on 8 nodes are considered here. In this case there is on each node a fraction of size $L \times (L/2)^3$ of the lattice, where the first factor refers to the time coordinate. The nodes operate in a SIMD mode and the corresponding sublattices are, therefore, updated simultaneously.

A first possiblity now is to update the link variables $U(x,\mu)$ in the order where the innermost loop runs over $x_0$, the next over $x_1$, then $x_2$, then $x_3$ and the last loop over the direction $\mu$. This is referred to as SF-updating, since it is the default order used in simulations of the Schrödinger functional.

Alternatively one may sweep through the lattice and update all link variables $U(x,\mu)$ at the current site $x$ before going to the next lattice point. This order has been used for the computation of $\bar{g}^2_{\rm TP}(L)$ and is, therefore, called TP-updating.

At $L/a = 24$ our program with SF-updating achieves a link update time of 34 $\mu$s (on each node). This number refers to the average update time in the mixture of heatbath and over-relaxation sweeps and includes the communication overhead. TP-updating is faster by about 25%. For comparision we mention that a similiar program, with the same random number generator, runs at a speed of about 3.1 $\mu$s per link on a CRAY-YMP.

Table 1. Relative variance $\mathcal{V}$ of $\partial S/\partial\eta$ at $\bar{g}^2_{\rm SF}(L) \simeq 3.6$

| $L/a$ | $\mathcal{V}$ | $L/a$ | $\mathcal{V}$ |
|---|---|---|---|
| 5 | 0.38 | 14 | 1.6 |
| 6 | 0.50 | 16 | 1.9 |
| 7 | 0.64 | 20 | 2.5 |
| 8 | 0.77 | 24 | 3.1 |
| 10 | 1.0 | 28 | 3.7 |
| 12 | 1.3 | 32 | 4.5 |

*3.3 Computational efficiency*

An important criterion for the choice of a particular finite volume renormalized coupling is the statistical precision that can be achieved in the numerical simulations. With this in mind we here discuss the variances and autocorrelation times of the two couplings introduced above.

The relative variance $\mathcal{V}$ of any observable $\mathcal{O}$ is defined by

$$\mathcal{V} = \left(< \mathcal{O}^2 > - < \mathcal{O} >^2\right) / < \mathcal{O} >^2 . \tag{3.12}$$

In table 1 the variance of $\partial S/\partial\eta$ is listed for an approximately fixed value of $\bar{g}^2_{\rm SF}(L)$. The behaviour of the variance is similiar at other values of the coupling. One notices a significant dependence on the lattice spacing roughly proportional to $(L/a)^{1.3}$.

In the case of the observable $\mathcal{O}_1$ [eq.(3.11)], the variance is nearly independent of the lattice spacing in the range $8 \leq L/a \leq 16$. Its value slightly increases from 1.3 at $\bar{g}^2_{\rm TP}(L) \simeq 8.1$ to 1.9 at $\bar{g}^2_{\rm TP}(L) \simeq 2.4$. The observable $\mathcal{O}_3$ has a much smaller variance than $\mathcal{O}_1$ and will not be considered any further, because the final error on the coupling is dominated by the error on the expectation value of $\mathcal{O}_1$.

Integrated autocorrelation times $\tau_{\rm int}$ were estimated by comparing the naive errors with our jacknife errors for large bin sizes. We used the Madras-Sokal formula [16], which approximates the required four-point autocorrelation function by its disconnected part, to calculate the statistical errors on the autocorrelation times. The truncation of the range over which the autocorrelation function is effectively summed is given by the bin length in our application. We found that this kind of error estimate is consistent in the cases where independent runs were carried out.

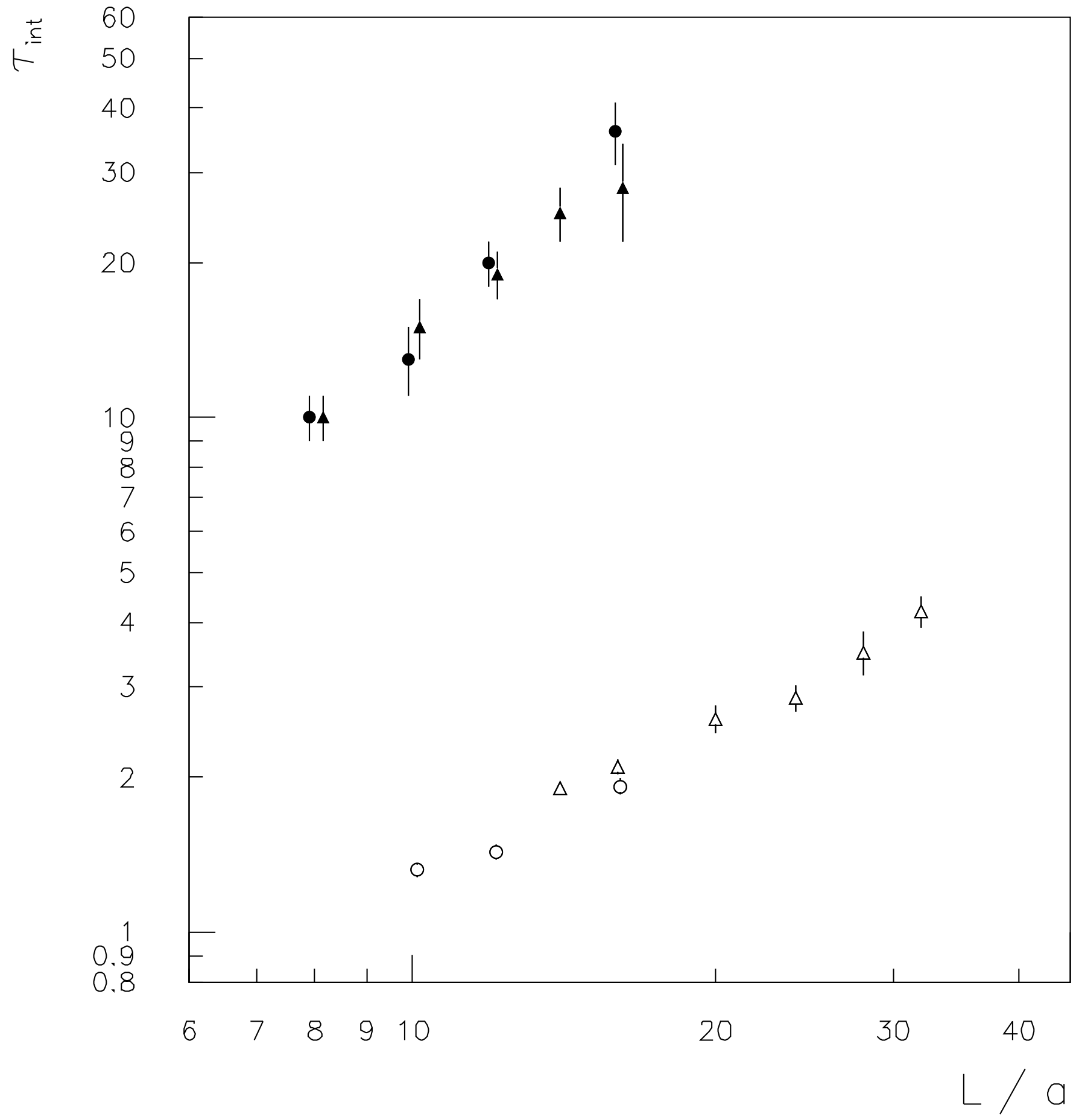

Fig. 1. Autocorrelation times of $\partial S/\partial\eta$ (open symbols) and $\mathcal{O}_1$ (filled symbols) for two different values of the renormalized coupling in each case.

In fig. 1 we show the integrated autocorrelation times $\tau_{\rm int}$ of $\partial S/\partial\eta$ (with SF-updating) and $\mathcal{O}_1$ (with TP-updating). We quote $\tau_{\rm int}$ in units of sweeps, not discriminating between over-relaxation and heatbath sweeps, since this unit is roughly proportional to the CPU-time. The figure shows that the autocorrelation times are rapidly growing, with no substantial dependence on the value of the renormalized coupling. In the present range of $L/a$ the data are described by $\tau_{\rm int} \propto (L/a)^z$ with $z \simeq 1.0$ and $z \simeq 1.8$ for $\partial S/\partial\eta$ and $\mathcal{O}_1$, respectively. It should be added here that the growth of $N_{\rm OR}$ with $L/a$ was chosen to be somewhat slower for TP-updating than for SF-updating (where

$N_{\rm OR}$ was always set to $L/2a$).

Taken together the computational effort to calculate the renormalized couplings $\bar{g}^2_{\rm SF}(L)$ and $\bar{g}^2_{\rm TP}(L)$ is in both cases scaling roughly proportionally to a power of $L/a$ with an exponent equal to 6 or slightly larger than 6. The prefactors are such that an order of magnitude less computer time is required to obtain $\bar{g}^2_{\rm SF}(L)$ to a given relative precision and for a given lattice size $L/a$.

After the computations had been finished we noted that SF-updating, when used to compute $\bar{g}^2_{\rm TP}(L)$, leads to a reduction of the autocorrelation time of $\mathcal{O}_1$ by a factor of 2. Even though the TP-updating program has a smaller link update time, it would thus have been profitable to use SF-updating in all cases. We have then also discovered that when TP-updating is used to simulate the Schrödinger functional, the autocorrelation time of $\partial S/\partial\eta$ increases by a factor 4 already on small lattices. Other updating schemes, such as checkerboard ordering, have also been tried, but none of them proved to be more efficient than SF-updating. At present we do not know what the cause of this unexpected behaviour is. The experience suggests, however, that it may be worthwhile to try simple variants of any given simulation algorithm before a large scale simulation is started.

## 4. Extrapolation to the continuum limit

We now proceed to discuss our results on the running couplings $\bar{g}^2_{\rm SF}(L)$ and $\bar{g}^2_{\rm TP}(L)$. Some of the data of refs.[5,9] will be included in the analysis. The new data are listed in tables 7–9 in appendix A.

Our principal aim in this section is to show that the extrapolation of a number of physical quantities to the continuum limit is well under control and that the resulting continuum amplitudes are universal, i.e. that they do not depend on the lattice action employed or on any other detail of the lattice definitions of the quantities considered.

Table 2. Expansion coefficients $\delta_0$ and $\delta_1$ for $\Sigma_{\rm TP}(2,u,a/L)$ [eq.(4.3)]

| $L/a$ | $\delta_0$ | $\delta_1$ |
|---|---|---|
| 4 | $-0.039219$ | $-0.014651$ |
| 5 | $-0.021339$ | $-0.007222$ |
| 6 | $-0.016852$ | $-0.004016$ |
| 7 | $-0.011382$ | $-0.002367$ |
| 8 | $-0.009344$ | $-0.001564$ |

### *4.1 Step scaling function*

On the lattice the step scaling function $\sigma(s,u)$ (subsect. 2.1) is approximated by a function $\Sigma(s,u,a/L)$ which describes the evolution of $\bar{g}^2(L)$ under changes of the scale $L$ by a factor $s$ at fixed $\beta = 4/g_0^2$ (a subscript SF or TP will be attached to $\Sigma(s,u,a/L)$ when the corresponding coupling is concerned). The defining equation is

$$\bar{g}^2(sL) = \Sigma(s,\bar{g}^2(L),a/L) \tag{4.1}$$

and we expect that

$$\sigma(s,u) = \lim_{a\to 0} \Sigma(s,u,a/L) \tag{4.2}$$

(if the continuum limit exists). Simulation results on $\Sigma(s,u,a/L)$ are available for $s=2$ and several values of $u=\bar{g}^2(L)$ and $a/L$. Our task is to extrapolate these data to the continuum limit.

In perturbation theory the cutoff dependence of the step scaling function can be studied by expanding

$$\Sigma(2,u,a/L)/\sigma(2,u) = 1 + \delta_0(a/L) + \delta_1(a/L)u + \delta_2(a/L)u^2 + \ldots \tag{4.3}$$

In the case of $\Sigma_{\rm SF}(2,u,a/L)$ the leading coefficient $\delta_0$ vanishes and the 1-loop and 2-loop coefficients $\delta_1$ and $\delta_2$ are tabulated in ref.[18]. The expansion coefficients of $\Sigma_{\rm TP}(2,u,a/L)$ have been computed to 1-loop order (table 2).

It should be noted at this point that $\Sigma_{\rm SF}(2,u,a/L)$ and the associated coefficients $\delta_k$ have an implicit dependence on the degree of improvement of the action. For the tree-level improved action $\delta_k$ is of order $a/L$ times a polynomial in $\ln(a/L)$ of degree $k-1$. 1-loop improvement implies a reduction of the degree by 1 and removes the terms of order $a/L$ from $\delta_1$ [which then becomes of order $(a/L)^2$]. And if one employs the 2-loop improved action, the order $a/L$ contributions to $\delta_2$ are cancelled in addition.

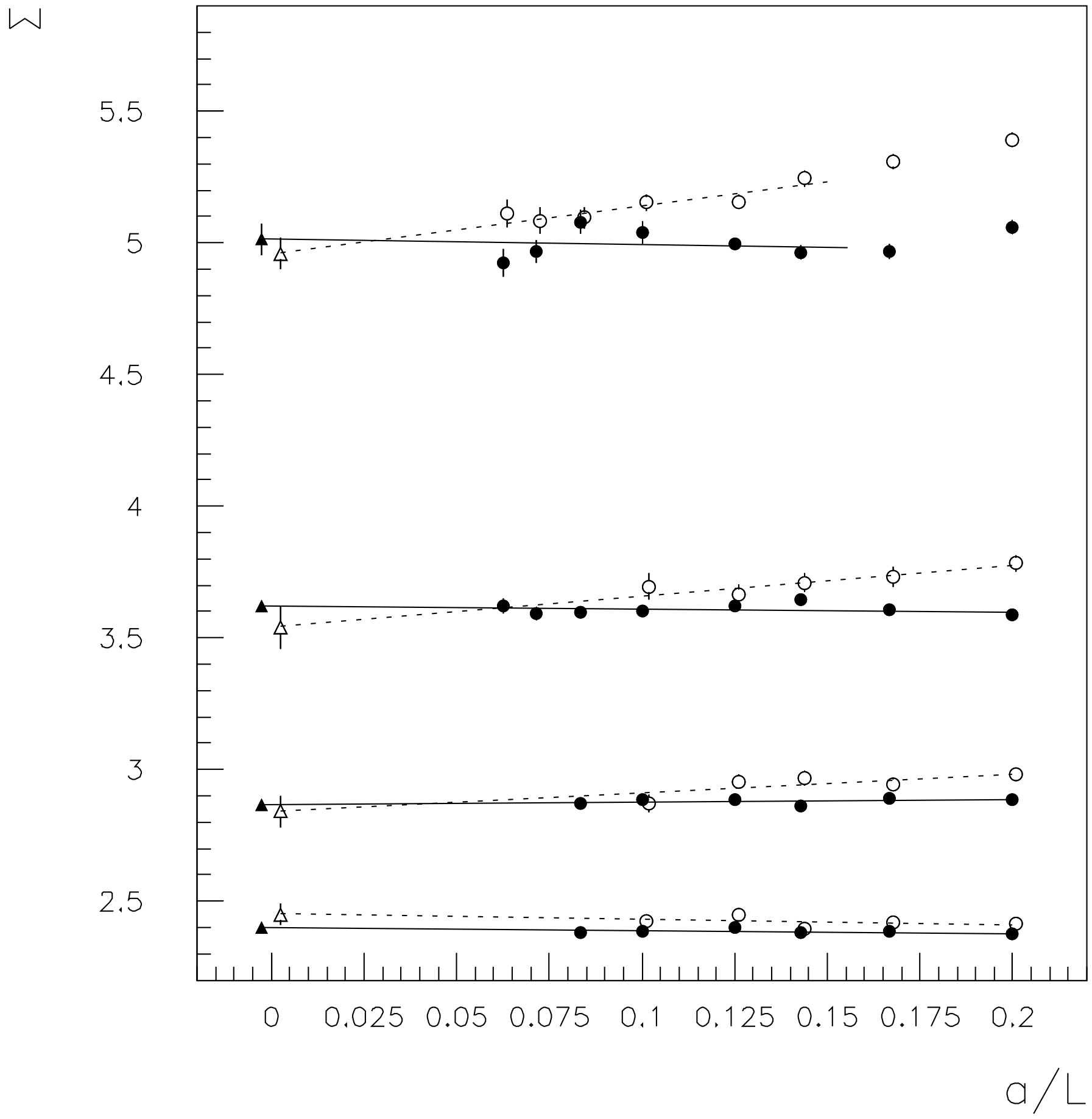

Fig. 2. Extrapolation of $\Sigma_{\rm SF}(2,u,a/L)$ to the continuum limit at $u=2.037$, 2.38, 2.84 and 3.55 (from bottom to top). The leftmost points are obtained by linear extrapolation and are slightly set apart for better readability. The action is tree-level improved (circles) or 1-loop improved (filled circles).

Perturbation theory thus suggests that in the case of $\Sigma_{\rm SF}(2,u,a/L)$ the leading cutoff effects are of order $a/L$. Moreover, the use of an improved action should speed up the convergence of $\Sigma_{\rm SF}(2,u,a/L)$ to the continuum limit. Our data are in fact compatible with a linear extrapolation in $a/L$ and one does observe that the slopes of the lines are reduced when going from the tree-level to the 1-loop improved action (see fig. 2). The plot also shows that the continuum limit is independent of the lattice action employed, thus confirming the expected universality of the step scaling function $\sigma_{\rm SF}(2,u)$.

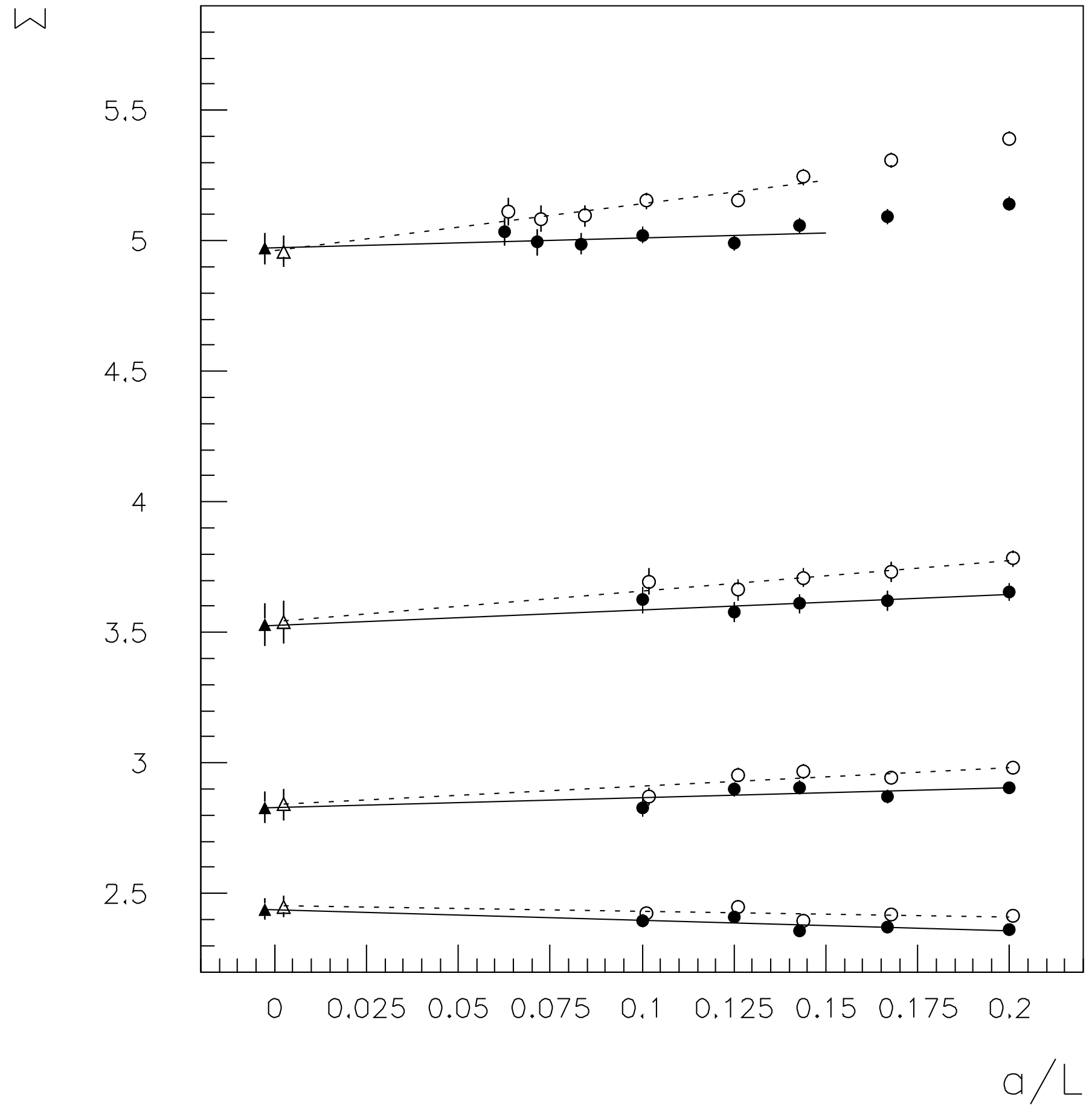

Fig. 3. Extrapolation of $\Sigma_{\rm SF}(2,u,a/L)$ (circles) and $\Sigma^{(2)}_{\rm SF}(2,u,a/L)$ (filled circles) to the continuum limit at the same couplings $u$ as in fig. 2. The simulations were done using the tree-level improved action.

Instead of $\Sigma_{\rm SF}(2,u,a/L)$ we may also try to extrapolate

$$\Sigma^{(2)}_{\rm SF}(2,u,a/L)=\frac{\Sigma_{\rm SF}(2,u,a/L)}{1+\delta_1(a/L)u+\delta_2(a/L)u^2}. \tag{4.4}$$

The continuum limits of $\Sigma_{\rm SF}(2,u,a/L)$ and $\Sigma^{(2)}_{\rm SF}(2,u,a/L)$ are trivially the same. We may, however, hope that the limit is reached faster in the case of $\Sigma^{(2)}_{\rm SF}(2,u,a/L)$, since the cutoff effects are here of order $u^3$. Note that this sort of improvement can be done with or without improving the action.

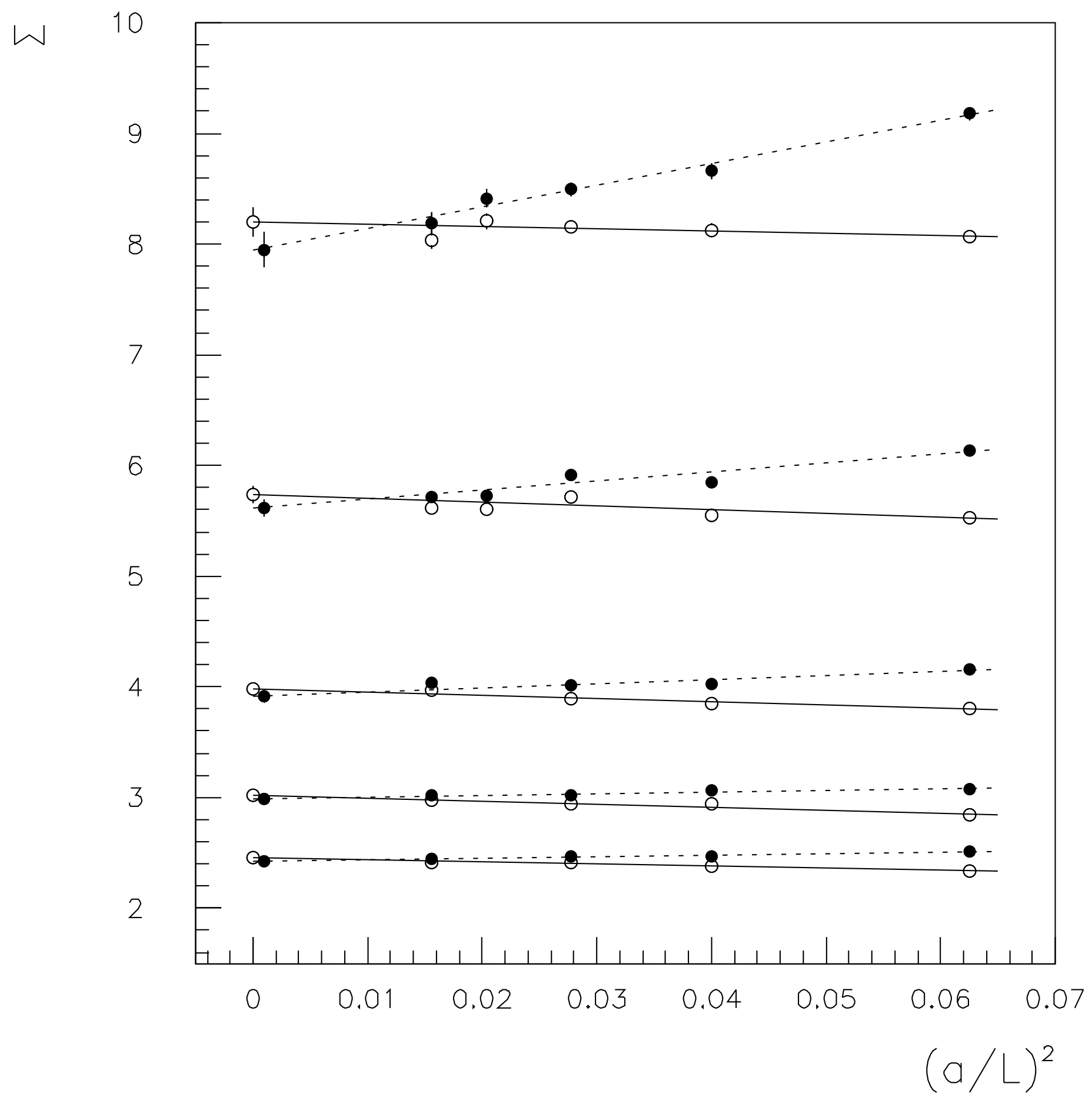

Fig. 4. Lattice spacing dependence of $\Sigma_{\rm TP}(2,u,a/L)$ (circles) and $\Sigma^{(1)}_{\rm TP}(2,u,a/L)$ (filled circles) at $u$ = 2.075, 2.430, 3.068, 4.108 and 5.597 (from bottom to top). The leftmost points are obtained by linear extrapolation.

The extrapolations of $\Sigma_{\rm SF}(2,u,a/L)$ and $\Sigma^{(2)}_{\rm SF}(2,u,a/L)$ (both calculated using the tree-level improved action) are compared in fig. 3. At the largest values of the coupling the improvement is evident directly from the figure, while the correction seems to go in the wrong direction at the lowest coupling $u$ = 2.037. This optical impression is, however, misleading. When the data

are fitted by

$$\Sigma^{(2)}_{\rm SF}(2,u,a/L) = \sigma_{\rm SF}(2,u) + \bar{d}(u)a/L, \tag{4.5}$$

the coefficients $\bar{d}(u)$ turn out to be zero within errors for all values of $u$. In the case of $\Sigma_{\rm SF}(2,u,a/L)$, on the other hand, the same fit yields slopes significantly different from zero for the three larger values of $u$. We conclude that perturbative improvement, i.e. passing from $\Sigma_{\rm SF}$ to $\Sigma^{(2)}_{\rm SF}$, is, in this case, very efficient in reducing the lattice artifacts.

In the case of $\Sigma_{\rm TP}(2,u,a/L)$ the cutoff effects are of order $(a/L)^2$ (cf. table 2). The data for $\Sigma_{\rm TP}(2,u,a/L)$ are in fact nearly independent of the lattice spacing and can easily be extrapolated to the continuum limit (see fig. 4). The "improved" quantity

$$\Sigma^{(1)}_{\rm TP}(2,u,a/L) = \frac{\Sigma_{\rm TP}(2,u,a/L)}{1+\delta_0(a/L)+\delta_1(a/L)u} \tag{4.6}$$

(which is also plotted in fig. 4) leads to a visible improvement only at the lowest two values of $u$. Significant cutoff effects are, however, observed at larger couplings, indicating that 2-loop or non-perturbative contributions are important for the description of lattice artifacts at values of $u$ greater than about 3. This is not really qualitatively different from the case of $\Sigma_{\rm SF}$, where the 2-loop terms in eq. (4.4) become important at such large values of the coupling. It is interesting to note in this connection that the cutoff effects in the two-dimensional non-linear $\sigma$–model are also not well described by perturbation theory unless the coupling is very small [3].

We conclude that perturbative improvement must be applied with caution. No general recommendation can be issued at this moment and simulations of sequences of lattices with decreasing spacings remain mandatory for a safe extrapolation to the continuum limit.

Our final results for the step scaling function $\sigma_{\rm SF}(2,u)$, together with a description of the extrapolation used in each case, are compiled in table 3. The last two lines in this table come from different sets of simulations and may be combined to yield $\sigma_{\rm SF}(2,3.55) = 4.98(4)$. It should be emphasized that all other extrapolations described above yield results compatible with the numbers quoted in table 3.

The values of $\sigma_{\rm TP}(2,u)$ used later in this paper to compute the evolution of $\bar{g}^2_{\rm TP}(L)$ have been obtained from $\Sigma_{\rm TP}(2,u,a/L)$ by linear extrapolation in $(a/L)^2$. The numbers have already been tabulated in ref.[9] and so are not reproduced here.

Table 3. Extrapolation of the step scaling function $\Sigma_{\rm SF}(2,u,a/L)$

| $u$ | action | quantity | $L/a-$range | $\sigma_{\rm SF}(2,u)$ | $\chi^2/$dof |
|---|---|---|---|---|---|
| 2.037 | 1-loop | $\Sigma_{\rm SF}^{(2)}$ | 5-12 | 2.384(14) | 2.5/4 |
| 2.380 | 1-loop | $\Sigma_{\rm SF}^{(2)}$ | 5-12 | 2.85(2) | 4.4/4 |
| 2.840 | 1-loop | $\Sigma_{\rm SF}^{(2)}$ | 5-16 | 3.59(2) | 4.6/6 |
| 3.550 | tree | $\Sigma_{\rm SF}$ | 7-16 | 4.97(6) | 2.8/4 |
| 3.550 | 1-loop | $\Sigma_{\rm SF}^{(2)}$ | 7-16 | 4.99(6) | 6.9/4 |

*4.2 Non-perturbative relation between $\alpha_{\rm SF}$ and $\alpha_{\rm TP}$*

In the continuum limit there exists a function $\phi(u)$ such that

$$\bar{g}^2_{\rm TP}(L)=\phi\left(\bar{g}^2_{\rm SF}(L)\right). \tag{4.7}$$

In particular, for small $u$ we infer from eqs.(2.14) and (2.19) that

$$\phi(u)=u+0.1195\times u^2+\ldots, \tag{4.8}$$

but $\phi(u)$ is also well-defined beyond perturbation theory. This function thus provides another opportunity to study the approach of the lattice data to the continuum limit.

On the lattice the relation between the couplings is of the form

$$\bar{g}^2_{\rm TP}(L)=\Phi\left(\bar{g}^2_{\rm SF}(L),a/L\right) \tag{4.9}$$

with some function $\Phi(u,a/L)$ which converges to $\phi(u)$ in the limit $a/L\to 0$. The leading cutoff effects are expected to be of order $a/L$. In table 4 we list the two renormalized couplings for a range of $L/a$, with the bare coupling adjusted in such a way that $\bar{g}^2_{\rm SF}(L)\simeq 2.0778$. The last column in this table is hence equal to $\Phi(u,a/L)$ at $u\simeq 2.0778$.

Within errors we do not in this case observe any cutoff effects beyond $L/a=6$. Again this is striking evidence for the existence of the continuum

Table 4. Cutoff dependence of $\bar{g}^2_{\rm TP}(L)$ at fixed $\bar{g}^2_{\rm SF}(L) \simeq 2.0778$

| $\beta$ | $L/a$ | $\bar{g}^2_{\rm SF}(L)$ | $\bar{g}^2_{\rm TP}(L)$ |
|---|---|---|---|
| 3.4057 | 4 | 2.0778(14) | 3.064(12) |
| 3.5504 | 6 | 2.0778(27) | 3.012(15) |
| 3.6585 | 8 | 2.0778(35) | 2.954(11) |
| 3.7464 | 10 | 2.0778(39) | 2.944(17) |
| 3.8150 | 12 | 2.0778(66) | 2.954(13) |
| 3.8709 | 14 | 2.0778(52) | 2.965(26) |
| 3.9200 | 16 | 2.0778(54) | 2.975(15) |

$\bar{g}^2_{\rm SF}(L)$ has been computed using the 1-loop improved action

limit. Extrapolation of the results for $L/a \geq 8$, allowing for a linear dependence on $a/L$, yields

$$\alpha_{\rm TP} = 0.2374(26) \quad \text{at} \quad \alpha_{\rm SF} = 0.16535. \tag{4.10}$$

We have here just considered one particular (and rather small) value of the coupling, but there is now little doubt that the continuum limit of $\Phi(u, a/L)$ is reached similiarly at other values of $u$, too.

*4.3 Low-energy scale*

So far we have been exclusively concerned with physical quantities defined in finite volume. We now introduce an infinite volume low-energy scale $r_0$ and compute the value of some reference box size $L_0^{\rm SF}$ in units of $r_0$. We then show that the ratio $L_0^{\rm SF}/r_0$ converges to a universal value in the continuum limit. This result is quite crucial as it allows us to relate the box sizes at which the renormalized couplings have been computed to the physical low-energy scales in the theory.

The scale $r_0$ is defined through the implicit equation

$$F(r_0) r_0^2 = 1.65, \tag{4.11}$$

where $F(r)$ denotes the force between heavy quarks at distance $r$ in infinite volume. The advantages of this particular definition and the numerical computation of $r_0$ have been discussed in ref.[17]. In nature $r_0$ is approximately equal to 0.5 fm, but it should be emphasized that $r_0$ is, in the present context, a purely theoretical reference scale with no precise phenomenological meaning.

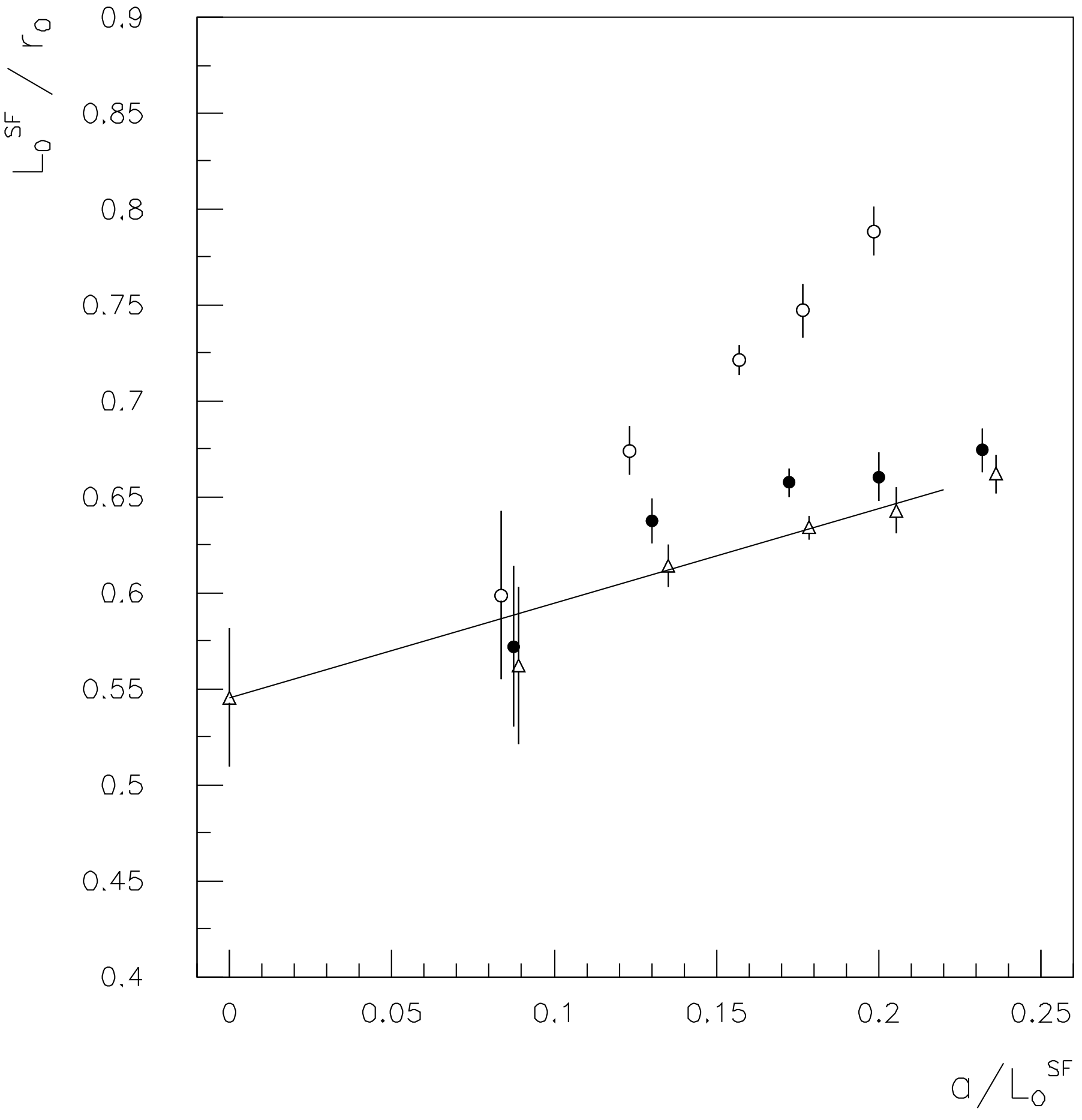

Fig. 5. Extrapolation of the ratio $L_0^{\mathrm{SF}}/r_0$ to the continuum limit. The data points are from simulations using the tree-level improved (circles), 1-loop improved (filled circles) and 2-loop improved (triangles) action. The leftmost point represents the result of a linear extrapolation of the data generated with the 2-loop improved action.

The box size $L_0^{\mathrm{SF}}$ has already been introduced in refs.[5,6]. It is implicitly defined by

$$\bar{g}_{\mathrm{SF}}^2(L_0^{\mathrm{SF}}) = 4.765. \tag{4.12}$$

To compute the ratio $L_0^{\mathrm{SF}}/r_0$ one needs to determine $L_0^{\mathrm{SF}}/a$ and $r_0/a$ at a given value of the bare coupling $\beta$. This has previously been done using the tree-level improved and 1-loop improved action [17]. It has already been observed there that the use of a 1-loop improved action reduces the lattice artifacts in the ratio $L_0^{\rm SF}/r_0$ significantly. We have now extended this analysis to the 2-loop improved action (cf. table 9 in appendix A).

Fig. 5 shows the lattice spacing dependence of the ratio $L_0^{\rm SF}/r_0$ for the three different actions. We again expect that the leading cutoff effects are of order $a/L$ and thus extrapolate the data linearly. Using the 1-loop improved action one obtains $L_0^{\rm SF}/r_0 = 0.573(37)$ in the continuum limit [17], while with the 2-loop improved action the result is

$$L_0^{\rm SF}/r_0 = 0.546(36). \qquad (4.13)$$

The data from the tree-level improved action are also consistent with this value, but are more difficult to extrapolate due to their stronger dependence on the lattice spacing. In any case fig. 5 represents further evidence for the universality of the continuum limit and underlines the necessity to study the cutoff dependence of the physical quantities of interest.

## 5. Applicability of perturbation theory

Perturbation theory is expected to apply at large momenta where the renormalized coupling is small. By comparing with our numerical results, we can now check to what extent this is true for the quantities considered.

### *5.1 Evolution of the running coupling*

Using the values of the step scaling function $\sigma_{\rm SF}(2,u)$ computed in sect. 4, the running coupling $\alpha_{\rm SF}(q)$ can be calculated at a sequence of momenta $q$ given in units of the reference box size $L_0^{\rm SF}$ (see table 5). The details of this calculation are as in ref.[5] and so are not discussed here again. It should however be emphasized that as a result of our use of the powerful APE computers the running coupling is now obtained with a statistical error of less than 1%.

Since the box sizes in table 5 range from about 0.27 fm to 0.017 fm (if we assume $r_0 = 0.5$ fm) it is perhaps not too surprising that the evolution of the coupling is rather well described by the 2-loop approximation to the Callan-Symanzik $\beta$–function [5]. More precisely if we solve the approximate

Table 5. $\alpha_{\rm SF}$ at scales given in units of $L_0^{\rm SF}$

| $qL_0^{\rm SF}$ | $\alpha_{\rm SF}(q)$ | 3-loop fit |
|---|---|---|
| 1.000 | 0.3792 | 0.3792 |
| 1.854 | 0.2825(16) | 0.2841(13) |
| 3.597 | 0.2260(14) | 0.2262(13) |
| 7.148 | 0.1894(14) | 0.1878(11) |
| 14.20 | 0.1621(13) | 0.1611(9) |

renormalization group equation

$$L\frac{\partial\bar{g}}{\partial L} = b_0\bar{g}^3 + b_1\bar{g}^5 + b_2^{\rm eff}\bar{g}^7 \tag{5.1}$$

with the initial condition $\bar{g}^2(L_0^{\rm SF}) = 4.765$ and choose

$$b_2^{\rm eff} = 0.35(12)/(4\pi)^3, \tag{5.2}$$

an excellent fit of the data is obtained (last column in table 5). The effective 3-loop term included in eq.(5.1) is rather small — its contribution decreases from about 7% at the largest coupling in table 5 to 1.4% at the smallest coupling.

The fit may be taken as an economic analytic representation of our data (including errors). In the following we shall use it to interpolate $\bar{g}_{\rm SF}^2(L)$ between the values listed in table 5. With little risk of running into uncontrolled extrapolation errors one may also rely on the fit to compute $\alpha_{\rm SF}(q)$ at momenta $q$ higher than those covered by the data.

We finally mention that the evolution of the other coupling $\bar{g}_{\rm TP}^2(L)$ is also well accounted for by the 2-loop $\beta$–function (cf. ref.[9]). The box sizes here range from approximately 0.2 fm to 0.006 fm, while the (relative) statistical errors are larger by as much as a factor of 2 compared to the errors quoted in table 5.

*5.2 Relation between $\alpha_{\rm SF}$ and $\alpha_{\rm TP}$*

The relation between our running couplings has already been discussed in subsect. 4.2. In particular, we have there been able to compute the value of $\alpha_{\rm TP}(q)$ non-perturbatively at a momentum $q$ implicitly determined by $\alpha_{\rm SF}(q) = 0.16535$. Considering table 5 this coupling is rather small and thus in the range where we should be able to compute $\alpha_{\rm TP}(q)$ using perturbation theory.

In the accessible range of couplings the 1-loop term in the series

$$\alpha_{\rm TP} = \alpha_{\rm SF} + 1.5017 \times (\alpha_{\rm SF})^2 + \ldots \tag{5.3}$$

[which one deduces from eqs.(2.14),(2.19)] is however quite large. As a consequence a straightforward application of the expansion is unsatisfactory. In particular, at $\alpha_{\rm SF} = 0.16535$ the right hand side of eq.(5.3) evaluates to 0.2064, thus only accounting for about half of the correction required to reproduce the numerical value $\alpha_{\rm TP} = 0.2374(26)$.

From the above we know that at large momenta the evolution of both couplings $\alpha_{\rm TP}$ and $\alpha_{\rm SF}$ is accurately described by the universal 2-loop approximation to the Callan-Symanzik $\beta$–function. A perturbative relation between $\alpha_{\rm TP}(q)$ and $\alpha_{\rm SF}(sq)$ (note the shifted scale) may, therefore, be expected to work well, provided a suitable scale factor $s$ is taken. For such considerations the precise choice of $s$ is not relevant, but only the range in which it lies. An intuitive choice to try is to set $s$ to the ratio between the $\Lambda$–parameters, viz.

$$s = \Lambda_{\rm SF}/\Lambda_{\rm TP} = 0.27620(2). \tag{5.4}$$

The perturbative relation then becomes

$$\alpha_{\rm TP}(q) = \alpha_{\rm SF}(sq) + {\rm O}\left([\alpha_{\rm SF}(sq)]^3\right), \tag{5.5}$$

i.e. the scale factor (5.4) has the effect of eliminating the 1-loop term.

So let us now evaluate eq.(5.5) at the momentum $q$ where $\alpha_{\rm SF} = 0.16535$. Using the effective 3-loop $\beta$–function discussed in subsect. 5.1, one finds

$$\alpha_{\rm SF}(sq) = 0.2289(6). \tag{5.6}$$

This is indeed close to the true value of $\alpha_{\rm TP}(q)$ given in eq.(4.10) and the remaining difference

$$[\alpha_{\rm TP}(q) - \alpha_{\rm SF}(sq)] \, / \, [\alpha_{\rm SF}(sq)]^3 = 0.7(3) \tag{5.7}$$

could well be accounted for by the higher order terms in eq.(5.5).

We thus conclude that perturbation theory correctly describes the relation between $\alpha_{\rm TP}$ and $\alpha_{\rm SF}$ at large momenta $q$, provided the matching is done at appropriatly shifted scales. We have shown that this is the case at a particular value of $q$, but since both couplings evolve with approximately the same $\beta$–function, the conclusion is bound to hold for a wider range of momenta.

*5.3 Relation between $\alpha_{\rm SF}$ and the bare coupling*

The bare perturbation expansion

$$\bar{g}_{\rm SF}^2(L) = g_0^2 + c_1(a/L)g_0^4 + c_2(a/L)g_0^6 + \ldots \tag{5.8}$$

has recently been worked out to 2-loop order [18] and we now would like to compare this result with our numerical data for the running coupling. In the continuum limit the bare coupling and the lattice spacing $a$ are simultaneously taken to zero in such a way that $a$ is an exponentially vanishing function of $g_0^2$. In this context it is hence consistent to drop all terms of order $a^p$, $p > 0$, contributing to the coefficients $c_k(a/L)$, which then reduce to polynomials in $\ln(L/a)$.

The resulting series is, however, of limited value, since it can only be applied if $g_0^2 \ln(L/a)$ is small. A similar situation is encountered in the continuum theory, when $\bar{g}_{\rm SF}^2(L)$ is expanded in powers of the coupling $\bar{g}_{\rm SF}^2(L')$ at scales $L'$ much smaller than $L$. Now if we choose $L'$ to be proportional to $a$, the two series may be combined to obtain an expansion of $\bar{g}_{\rm SF}^2(L')$ in powers of the bare coupling with no large logarithms. In terms of $\alpha_0 = g_0^2/4\pi$ the general form of this expansion is

$$\alpha_{\rm SF}(s/a) = \alpha_0 + d_1(s)\alpha_0^2 + d_2(s)\alpha_0^3 + \ldots \tag{5.9}$$

where $s$ is a scale factor to be chosen with care. It should be emphasized that in this equation $\alpha_{\rm SF}(s/a)$ is defined in the continuum theory. In particular, the momentum $s/a$ must be given in units of some physical reference scale, such as $L_0^{\rm SF}$ or $r_0$, for the relation to be meaningful.

To make practical use of the series (5.9) the scale factor $s$ should be chosen such that the low-order coefficients $d_k(s)$ are reasonably small (as far as possible). Different ways to achieve this have been discussed in refs.[18,19]. As already noted above the precise value of $s$ is unimportant, since small differences will be compensated by the associated changes of the 1- and 2-loop terms in eq.(5.9). As in the case of the relation between $\alpha_{\rm SF}$ and $\alpha_{\rm TP}$ an obvious possibility is to adjust $s$ so that the 1-loop term in the expansion vanishes. This leads to the series

$$\alpha_{\rm SF}(8.83/a) = \alpha_0 + 1.287 \times \alpha_0^3 + {\rm O}(\alpha_0^4). \tag{5.10}$$

Table 6. Estimates of $\alpha_{\rm SF}(q)$ at $q = 10/a(2.85)$

| $\alpha_{\rm SF}$ | method |
|---|---|
| 0.1098 | 1-loop in $\alpha_0$ |
| 0.1115 | 2-loop in $\alpha_0$ |
| 0.1110 | 1-loop in $\tilde{\alpha}_0$ |
| 0.1128 | 2-loop in $\tilde{\alpha}_0$ |

Note that with this choice of $s$ the 2-loop term is only 1–2% of the leading term in the interesting range of bare couplings.

The unexpectedly large scale factor $s$ appearing in eq.(5.10) has long been a source of worry and was interpreted as a sign for $\alpha_0$ to be a "bad" expansion parameter [19,20]. It was then suggested to replace $\alpha_0$ by an "improved" bare coupling such as

$$\tilde{\alpha}_0 = \alpha_0/P, \tag{5.11}$$

for example, where $P$ is the average plaquette (in infinite volume) belonging to the given value of $g_0^2$ [20]. The associated expansion of $\alpha_{\rm SF}$,

$$\alpha_{\rm SF}(1.17/a) = \tilde{\alpha}_0 + 0.951 \times \tilde{\alpha}_0^3 + {\rm O}(\tilde{\alpha}_0^4), \tag{5.12}$$

then comes with a more comfortable scale factor $s$ (which was again fixed by requiring the 1-loop term to vanish).

We now evaluate these formulae at bare coupling $\beta = 2.85$. The associated lattice spacing is denoted by $a(2.85)$. It is fairly small — the corresponding momentum $1/a(2.85)$ is about 8 GeV (if we assume $r_0 = 0.5$ fm) — and so we expect the perturbation expansion to apply.

From eqs.(5.10) and (5.12) the coupling $\alpha_{\rm SF}(q)$ is obtained at two values of the momentum $q$ given in units of $a(2.85)$. Due to the contracting nature of the evolution, the relative errors at different scales are not directly comparable. We hence evolve all results for $\alpha_{\rm SF}$ to the reference energy $q = 10/a(2.85)$. This is done using the 3-loop fit of the evolution of $\alpha_{\rm SF}$ discussed in subsect. 5.1.

The error quoted in eq.(5.2) is negligible here and the results are listed in table 6.

To compare with the non-perturbative values of $\alpha_{\rm SF}$ (which were determined at scales given in units of $L_0^{\rm SF}$) we need to compute the conversion factor $a(2.85)/L_0^{\rm SF}$. This can be done using the tree-level, 1-loop or 2-loop improved action. The required information can be found in refs.[5,17] and in table 9. As a result one obtains $a(2.85)/L_0^{\rm SF} = 0.0834(5)$, $0.0874(5)$ and $0.0889(5)$ in the three cases. We take the last value and conservatively include 0.0015 as a systematic error on the scale ratio. Note however the trivial fact that a continuum limit cannot be taken here in contrast to all scale ratios appearing within the finite-size method.

Using the 3-loop fit of our data the numerical value of the SF-coupling at $q = 10/a(2.85)$ can now be quoted as

$$\alpha_{\rm SF}(q) = 0.1135(8), \tag{5.13}$$

where the error is obtained by taking the (absolute) sum of the statistical error and the systematic error from the scale ratio.

We conclude that the 2-loop formula (5.12) reproduces the numerical value of $\alpha_{\rm SF}$ within 1 error margin. When using the standard bare coupling as an expansion parameter, the 2-loop result is off by about 1.8% (2.5 error margins). Both series show no pathology if the scale factor $s$ is determined by requiring the 1-loop term to vanish. In particular, without independent numerical control it would seem difficult to us to argue that one of the expansions is better than the other.

## 6. Computation of $\alpha_{\overline{\rm MS}}$

We finally discuss the computation of $\alpha_{\overline{\rm MS}}(q)$ at large momenta $q$ given in units of the physical low-energy scale $r_0$ (subsect. 4.3). The conversion to the $\overline{\rm MS}$ scheme is of conceptual importance, because all reference to a finite volume disappears in this step and one gets a result, which is directly relevant to the theory in infinite volume.

The computation is straightforward. Through the 3-loop fit of our numerical results, described in subsect. 5.1, the coupling $\alpha_{\rm SF}(q)$ is available at all momenta $q \geq 1/L_0^{\rm SF}$ given in units of $1/L_0^{\rm SF}$. Using the conversion factor

(4.13) we can easily pass to units of $1/r_0$. Then, at large momenta $q$, the 1-loop formula

$$\alpha_{\overline{\rm MS}}(q) = \alpha_{\rm SF}(sq) + {\rm O}([\alpha_{\rm SF}(sq)]^3) \tag{6.1}$$

is employed, with the scale factor $s$ given by

$$s = \Lambda_{\rm SF}/\Lambda_{\overline{\rm MS}} = 0.44567. \tag{6.2}$$

This choice is motivated by our experience with the perturbation expansions discussed in the preceeding section.

To illustrate the procedure, we choose $q = 20/r_0$ which is roughly equal to the momentum $1/a(2.85)$ considered above. One then obtains

$$\alpha_{\overline{\rm MS}}(q) = 0.2075(39)(89) \tag{6.3}$$

The second error here is equal to $\alpha_{\rm SF}(sq)^3$ and is quoted to remind us of our ignorance of the 2-loop term in the perturbative relation (6.1) between $\alpha_{\overline{\rm MS}}$ and $\alpha_{\rm SF}$. The associated coefficient is currently being computed and will be known soon [18,21] so that this source of error will be better under control. The first error in eq.(6.3) is obtained by combining the statistical errors from the scale conversion factor $L_0^{\rm SF}/r_0$ and the evolution of $\alpha_{\rm SF}$ in quadrature.

A more precise result on $\alpha_{\overline{\rm MS}}(q)$ is obtained at larger momenta, because the coupling becomes smaller and has a weaker scale dependence. At $q = 200/r_0$ for example (which corresponds to about 80 GeV in physical units), one gets

$$\alpha_{\overline{\rm MS}}(q) = 0.1288(15)(21) \tag{6.4}$$

and thus a total error of only 3% (if we assume the perturbative corrections to eq.(6.1) are no larger than quoted).

We would like to emphasize that in these calculations we do not need to assume that $\alpha_{\overline{\rm MS}}$ evolves according to perturbation theory. The conversion from the SF to the $\overline{\rm MS}$ scheme, using eq.(6.1), takes place in the last step of the computation, at the momentum where the $\overline{\rm MS}$ coupling is desired. The evolution of $\alpha_{\rm SF}$, on the other hand, is numerically controlled.

## 7. Conclusions

While our results do not prove that the continuum limit of the SU(2) lattice gauge theory exists in the expected way, they represent further impressive evidence for this to be true. All the numerical data are perfectly compatible with the assumption, at an unprecedented level of precision. To see this it was however important to study the renormalized couplings (and the other quantities of interest) on sequences of lattices with decreasing lattice spacings. The necessity of such an expensive procedure is made particularly clear by fig. 5, where we discuss the relation between physical scales defined in finite and infinite volume.

With our recursive finite-size technique we are able to follow the running couplings from low to rather high momenta. As already discussed in our previous work [3–9], the evolution of $\alpha_{\rm SF}$ and $\alpha_{\rm TP}$ is well described by perturbation theory in the whole range of momenta covered. The perturbative relation between the couplings must however be applied with care. Our experience is that it is important to do the matching at appropriately shifted scales. A good choice of scale factor is the corresponding ratio of $\Lambda$–parameters.

The same comment also applies when considering the expansion of $\alpha_{\rm SF}$ in powers of the bare coupling. Here we have observed that the use of an "improved" bare coupling $\tilde{\alpha}_0$ [eq.(5.11)] leads to slightly better results, although there is no sign from the series alone (which is known to 2-loop order) that the standard bare coupling $\alpha_0$ would be a bad choice of expansion parameter at the appropriate scale.

We finally note that at the level of precision reached, the perturbative formulae employed to convert from one coupling to another must be worked out to 2-loop order. In particular, the error on our results for $\alpha_{\overline{\rm MS}}$ at large momenta is now dominated by the truncation at 1-loop order of the expansion of this coupling in powers of $\alpha_{\rm SF}$.

Table 7. Pairs of running couplings at fixed bare coupling (tree-level improved action)

| $\beta$ | $L/a$ | $\bar{g}^2_{\rm SF}(L)$ | $\bar{g}^2_{\rm SF}(2L)$ |
|---|---|---|---|
| 2.7170 | 5 | 3.5500(59) | 5.364(28) |
| 2.7903 | 6 | 3.5500(66) | 5.320(31) |
| 2.8610 | 7 | 3.5500(67) | 5.230(31) |
| 2.9158 | 8 | 3.5500(75) | 5.151(27) |
| 3.0091 | 10 | 3.5500(77) | 5.160(31) |
| 3.0767 | 12 | 3.5500(81) | 5.095(36) |
| 3.1401 | 14 | 3.5500(70) | 5.089(49) |
| 3.1874 | 16 | 3.5500(80) | 5.111(51) |

## Appendix A

We here collect our new data obtained from simulations of the Schrödinger functional on the APE computers. We were first interested to improve on the accuracy of the old data for the step scaling function $\Sigma_{\rm SF}(2,u,a/L)$ at $u = 3.55$, the largest coupling considered, using the tree-level improved action (table 7).

We have then computed the step scaling function $\Sigma_{\rm SF}(2,u,a/L)$ at the same values of $u$ previously considered, but now using the 1-loop improved action. These results are collected in table 8.

Finally the existing data to determine the conversion factor $L_0^{\rm SF}/r_0$ [5,17] were complemented by computing the $\beta$ values at which $\bar{g}^2_{\rm SF}(L) = 4.765$ as a function of the lattice size $L/a$ for the case of the 2-loop improved action (table 9).

Table 8. Pairs of running couplings at fixed bare coupling (1-loop improved action)

| $\beta$ | $L/a$ | $\bar{g}^2_{\rm SF}(L)$ | $\bar{g}^2_{\rm SF}(2L)$ |
|---|---|---|---|
| 3.4448 | 4 | 2.0370(22) | 2.366(8) |
| 3.5207 | 5 | 2.0370(20) | 2.375(6) |
| 3.5880 | 6 | 2.0370(20) | 2.386(9) |
| 3.6463 | 7 | 2.0370(21) | 2.383(9) |
| 3.6984 | 8 | 2.0370(22) | 2.401(10) |
| 3.7840 | 10 | 2.0370(21) | 2.384(11) |
| 3.8506 | 12 | 2.0370(28) | 2.383(11) |
| 3.1775 | 4 | 2.3800(30) | 2.869(9) |
| 3.2576 | 5 | 2.3800(27) | 2.886(9) |
| 3.3201 | 6 | 2.3800(26) | 2.890(12) |
| 3.3852 | 7 | 2.3800(29) | 2.860(12) |
| 3.4342 | 8 | 2.3800(30) | 2.888(13) |
| 3.5205 | 10 | 2.3800(26) | 2.888(14) |
| 3.5934 | 12 | 2.3800(29) | 2.870(14) |
| 3.0151 | 5 | 2.8400(27) | 3.589(10) |
| 3.0797 | 6 | 2.8400(32) | 3.607(14) |
| 3.1343 | 7 | 2.8400(28) | 3.645(21) |
| 3.1900 | 8 | 2.8400(37) | 3.622(21) |
| 3.2759 | 10 | 2.8400(35) | 3.602(17) |
| 3.3423 | 12 | 2.8400(40) | 3.598(20) |
| 3.4035 | 14 | 2.8400(50) | 3.592(24) |
| 3.4556 | 16 | 2.8400(48) | 3.622(26) |
| 2.7700 | 5 | 3.5500(46) | 5.060(24) |
| 2.8355 | 6 | 3.5500(51) | 4.967(25) |
| 2.8946 | 7 | 3.5500(43) | 4.964(25) |
| 2.9442 | 8 | 3.5500(54) | 4.994(20) |
| 3.0257 | 10 | 3.5500(58) | 5.038(42) |
| 3.0974 | 12 | 3.5500(86) | 5.080(44) |
| 3.1579 | 14 | 3.5500(79) | 4.966(41) |
| 3.2048 | 16 | 3.5500(82) | 4.924(50) |

Table 9. Bare couplings vs. lattice size at $\bar{g}^2_{\rm SF}(L) = 4.765$ (2-loop improved action)

| $L/a$ | $\beta$ | $L/a$ | $\beta$ |
|---|---|---|---|
| 5 | 2.5600(17) | 8 | 2.7279(27) |
| 6 | 2.6256(25) | 10 | 2.8061(24) |
| 7 | 2.6807(23) | 12 | 2.8756(30) |